\documentclass[12pt,draftcls,journal,onecolumn,romanappendices]{IEEEtran}

\usepackage[dvips]{hyperref}
\usepackage[dvips]{graphicx}
\usepackage{float,makeidx,cite}
\usepackage{amssymb,amsmath,amsthm}

\providecommand{\abs}[1]{\lvert#1\rvert}
\newcommand{\ud}{\,\mathrm{d}}

\newtheorem{Theoi1}{Theorem}
\newtheorem{Theoi2}[Theoi1]{Theorem}

\title{Cooperative Relaying for Large Random Multihop Networks}
\author{Amogh Rajanna and Mos Kaveh, \IEEEmembership{Life Fellow, IEEE}}
\begin{document}
\maketitle
\begin{abstract}
In this paper, we propose a new relaying protocol for large multihop networks combining the concepts of cooperative diversity and opportunistic relaying. The cooperative relaying protocol is based on two diversity mechanisms, incremental redundancy combining and repetition combining. We assume that nodes in the large multihop network are modeled by a homogeneous Poisson Point Process and operate under Rayleigh fading and constant power transmission per node. The performance of the proposed relaying protocol is evaluated through the progress rate density (PRD) of the multihop network and compared to the conventional multihop relaying with no cooperation. We develop an analytic approximation to the PRD based on the concept of decoding cells. The protocol parameters are optimized to maximize the PRD of network. We show that the cooperative relaying protocol provides significant throughput improvements over conventional relaying with no cooperation in a large multihop network.
It is also shown that incremental redundancy combining provides a higher gain in PRD relative to repetition combining. The gain in PRD has near constant value at all values of the path loss exponent and is monotonic in diversity order.
\end{abstract}

\section{Introduction}
\label{new_intro}
Large multihop networks are characterized by source and destination pairs spread over a wide area separated by a long distance with large number of relay nodes in between them. In such networks, the data packets can be efficiently transported from source to destination by employing opportunistic relaying protocols. In opportunistic relaying, one relay is selected from among potentially multiple relay nodes to forward data packets. In the literature, many versions of opportunistic relaying have been proposed with the key difference being the criterion used for selecting one relay from among multiple candidates. For example, one option would be to choose the relay node which is farthest from the source for forwarding \cite{Baccelli}. A similar criterion would be to choose the relay which is closest to the destination \cite{Zhao}. A criterion based on the channel state information was proposed in \cite{Ganti} and \cite{Anghel} where the relay with the best channel to the destination is selected.

In \cite{Baccelli}, an opportunistic relaying protocol for a large multihop network was proposed, where in every hop the relay farthest from the source is selected. A fully distributed implementation of the relay selection procedure was proposed. The protocol can communicate packets from any two nodes in the network via multiple hops without any prior connectivity. The work of \cite{Blomer_II} proposes to study the relaying protocol of \cite{Baccelli} by defining the progress rate density of the network and optimize the protocol to maximize the network throughput.
The authors of \cite{Ganti} study the performance of opportunistic relaying on the downlink of a two hop cellular system. The criterion for relay selection was having the best channel to the destination. The performance gain of two hop communication without cooperative diversity relative to direct communication between source and destination was quantified rigorously.

Cooperative communication in a single relay setting where the source and single relay transmit information to the destination has been studied thoroughly \cite{JNL}. The source and relay transmit orthogonally in time to the destination exploiting space and time diversity leading to improvements in throughput and outage. The performance gain obtained depends on the specific type of diversity technique employed by source and single relay. One type involves the source and relay transmitting to the destination using a distributed space time code. Cooperative diversity in the context of multiple relays assisting the source in a two hop communication to the destination has been extended in a similar way\cite{JNL}.

Cooperative diversity techniques developed have been applied exclusively to study source destination communication via two hops.
More research effort is necessary to understand the performance of cooperative diversity protocols in networks where source destination communicate via more than two hops and employ opportunistic relaying. Although the number of hops between source and destination is dictated by the size of the network, the network performance is expected to improve when cooperative diversity is present within the framework of opportunistic relaying since the best relay is chosen to perform cooperative transmission. For example, \cite{Andre} provides experimental data illustrating the gains in network performance when cooperative diversity is employed in a multihop network verifying the results reported in \cite{Hasna}. It is shown that the end to end success probability is nearly $1$ with smaller average delays when relays cooperatively transmit in a source to destination communication. In \cite{Mehta}, the performance improvements in a multihop network when relays combine cooperative diversity and packet buffering is quantified. The works of \cite{Urgaonkar,Draper} provide algorithms for resource allocation problems in multihop communication from source to destination when relays cooperate and accumulate mutual information leading to substantial improvements in energy efficiency and delay. 

In \cite{Zhao}, cooperative diversity is applied to a scenario where source destination communicate over a line network of finite length with fixed number of equidistant relays in between them. The cooperative diversity scheme employed is incremental redundancy combining, where the codeword of a data packet is split into non overlapping blocks via puncturing and are transmitted incrementally by the source and relays to the destination. It is shown that such a scheme has improvement in throughput, outage and energy efficiency over conventional relaying protocol with no cooperation.

The focus of this paper is on large multihop networks where source and destination typically communicate by more than two hops. The goal is to investigate the performance of incremental redundancy combining as a cooperative diversity scheme in a large multihop network with opportunistic relaying. The key difference between the present work and \cite{Zhao} stems from the network topology and wireless channel model. While \cite{Zhao} studies a line network of finite length with equidistant relays between source and destination and the wireless channel is affected by fading only, this paper studies a large multihop network and the the wireless channel has fading, path loss and interference.

In this paper\footnote{A part of the results presented in this paper appear in \cite{amogh_c1}.}, we model the nodes of a large multihop network by a homogeneous Poisson Point Process (PPP) \cite{Andrews_I}. In such a network, we propose to use incremental redundancy combining and repetition combining as cooperative diversity methods in conjunction with opportunistic relaying for multihop communication between source and destination. The resulting protocol is termed \emph{cooperative relaying protocol}. The performance of cooperative relaying protocol is quantified through progress rate density (PRD) of the network. The parameters of the protocol are optimized based on an analytic approximation to the PRD. The main results of the paper are
\begin{itemize}
\item The cooperative relaying protocol provides gains in throughput compared to conventional relaying with no cooperation since the relay nodes combine different transmissions for decoding one information packet thus extracting spatial and time diversity inherent in the network.

\item Two forms of combining are studied in the paper, incremental redundancy combining and repetition combining, a form of diversity where the relay nodes perform maximal ratio combining of the transmissions from source and previous relays. Since incremental redundancy combining provides new parity symbols in every relay transmission in addition to the space and time diversity, it achieves a larger gain in throughput relative to repetition combining.

\item The gain in PRD due to cooperative relaying protocol is monotonic in the diversity order $M$. In this paper, diversity order is defined as the number of diverse transmissions that a relay node combines to decode one information packet. For example, in the case of incremental redundancy combining the gain in PRD from $M=1$ to $M=2$ is $26.5\%$ whereas from $M=2$ to $M=3$, the gain is $9.3\%$.

\item The cooperative relaying protocol achieves a near constant gain in PRD as a function of $\alpha$, path loss exponent. For example, incremental redundancy combining with $M=2$ provides a gain of $26.5\%$ and $23.5\%$ at $\alpha=3$ and $\alpha=4$ respectively.

\end{itemize}
The results of the paper emphasize the potential for throughput improvements in a large multihop network by employing either incremental redundancy combining or repetition combining mechanisms over conventional relaying protocols with no cooperation.

The organization of the rest of the paper is outlined below. Section \ref{Multihop_relay} describes the cooperative relaying protocol for a large multihop network. In Section \ref{sysmod}, the system model and assumptions are presented. Section \ref{appro_prd} develops the analytic approximation to the PRD metric. In Section \ref{opti_prd}, a discussion on the protocol optimization is presented. Section \ref{hig_CooRe} proposes improvements to the cooperative relaying protocol. Section \ref{sim_res} presents the numerical results of the paper. Section \ref{con_cl} contains the conclusions.
\section{Cooperative Relaying protocol for Multihop Network}
\label{Multihop_relay}
In this section, we present the basic cooperative relaying protocol for a multihop random network. We assume a large multihop random network where nodes have access to a common bandwidth resource i.e., frequency band and time. Information packets are communicated from each source node to its destination node via multiple hops with the aid of relays. A key assumption is that the destination is at a large random distance from the source. Since the destination is at a random distance from the source, there is no predefined multihop path between them. It is assumed that the hops are isotropic. Examples of such networks include sensor networks reporting measurements to a central node, military network in the battlefield and mobile networks with user mobility.

The source node generates a codeword for an information packet and transmits it to the destination via a multihop version of incremental redundancy combining which is explained in the following. The codeword is split into non overlapping blocks by a puncturing process. The source transmits the $1^{st}$ block of the codeword which is received by potential relay nodes. One relay node is chosen to transmit the $2^{nd}$ block of the codeword and the procedure for relay selection is explained below.

\subsection{Relay Selection Procedure}
\label{Re_Se}
Even though all potential relay nodes receive the $1^{st}$ block of the codeword from the source, based on the instantaneous SINR conditions only a fraction of them will be able to decode the data packet. Each relay node which decodes the data packet has an associated \textit{progress}. The progress of a relay is defined as \textit{the distance from the source in the source-destination direction over which the information bits are communicated}. It is assumed that the data packet contains information about source and destination locations and also every node in the network knows its own location. So the relay nodes which successfully decode the data packet will be able to compute the progress they offer. The relays then participate in a distributed contention scheme to select the relay which offers the maximum progress as the forwarding relay $1$ to transmit the $2^{nd}$ block of the codeword.

The contention scheme for distributed relay selection is based on the one proposed in \cite{Baccelli}.
Each of the relays encode the progress they offer into a $P$ bit vector $b_1b_2\cdots b_P$. The value of $P$ is chosen a priori and depends on the network dimensions. The relays then participate in a contention period of duration $P$ time units. Each of the $P$ bits in the bit vector exclusively determine the activity in each of the $P$ time units.
For every $0$ bit in the bit vector, the relay listens to the channel during the corresponding time unit and for every $1$ bit in the bit vector, the relay transmits a pulse into the channel. Each relay starts its contention activity with the bit $b_P$ and proceeds all the way to bit $b_1$. For example, if a relay has the following bit vector $000110$, then it listens to the channel for first three time units, transmits two consecutive pulses and again listens to the channel in the last time unit. During a listening period, if a relay detects a pulse in the channel then it quits the relay selection process since it knows that another relay has a larger progress. In this way, the only relay that survives the contention period is the one with the most progress from the source\footnote{Although a single relay is capable of decoding multiple packets, it is less likely that the same relay node will offer maximum progress for the multiple packets and thus be the forwarding relay for all of them. In any case, we assume that even if a relay node has multiple packets for forwarding, it prioritizes them based on the ascending order of progress of packets and time spent by the packets in the queue etc.}. It is also assumed that the source node listens to the contention period. If it detects that no relay has been selected for forwarding the packet, then it retransmits the $1^{st}$ block of the codeword and the selection procedure repeats.

\subsection{Incremental Redundancy Combining}
\label{IR_com} 
When the forwarding relay $1$ transmits the $2^{nd}$ block of the codeword, all the potential relay nodes (and also the destination) combine the received $2^{nd}$ block of the codeword with the previously received $1^{st}$ block of the codeword from the source and try to decode the data packet. Based on the instantaneous SINR conditions, some fraction of the potential relay nodes will be able to decode the data packet by combining the $1^{st}$ and $2^{nd}$ blocks of the codeword. Subsequently, they participate in the distributed contention scheme of section \ref{Re_Se} to select the forwarding relay $2$.

Which block of the codeword does the forwarding relay $2$ transmit? The answer depends on the number of blocks of the codeword the relay nodes in the network (also destination) are allowed to combine for decoding. If the relay nodes are allowed to combine two blocks of the codeword for decoding\footnote{The number of blocks of the codeword the relay nodes are allowed to combine for decoding is the same as the number of blocks the codeword is split into by the source node initially.}, then every forwarding relay transmits a block of the codeword which is complementary to the most recent block it has received. For example, since the forwarding relay 2 receives the $2^{nd}$ block of the codeword from the previous forwarding relay, it transmits the $1^{st}$ block of the codeword. Similarly the forwarding relay $5$ receives the $1^{st}$ block of the codeword from forwarding relay $4$ and hence transmits the $2^{nd}$ block of the codeword.

This process of cooperative relaying whereby the relay nodes combine two blocks of the codeword for decoding, one from current forwarding relay and another from previous forwarding relay, and distributively select a forwarding relay which transmits an alternating block of the codeword continues until the data packet reaches the destination and is decoded successfully.

In a similar manner, a generalized version of incremental redundancy combining for multihop can be extended. Assume that the relay nodes are allowed to combine $M>2$ blocks of the codeword for decoding. In such a case, every time a forwarding relay transmits a current block of the codeword, all the potential relay nodes combine the current block of the codeword with all the $M-1$ recently received blocks of the codeword and make an attempt to decode the data packet. The relay nodes which decode the data packet participate in a distributed contention scheme to select the forwarding relay. The forwarding relay transmits a block of the codeword which is complementary to the $M-1$ recently received blocks of the codeword. For example if $M=3$, the first $M$ blocks of the codeword are transmitted by  source and forwarding relays $1$ and $2$ respectively. From forwarding relay $3$ onwards, the transmitted block of the codeword is complementary to the two most recent blocks received. In other words, forwarding relay $i$ transmits $\left(q(i)+1\right)^{th}$ block of the codeword, where $q(i)=\mod(i,3)$. For example, forwarding relays $4$ and $8$ transmit the $2^{nd}$ and $3^{rd}$ blocks of the codeword respectively. This process of cooperative relaying continues until the data packet reaches the destination and an ACK is sent back after successful decoding.

\subsection{Performance Metric}
\label{per_met}
Since the source and destination are separated by a random distance with no predefined path, the performance metric introduced in \cite{Andrews_II} for multihop networks namely the Random Access Transport Capacity (RATC), which accounts for the number of hops and thus the time delay in transporting information bits from source to destination is not suitable for the network system model in this paper\footnote{RATC is more suited to the system model where source and destination communicate with a fixed number of equidistant relays in between them \cite{Andrews_II, Vaze_I}.}. For the network model in our paper, the key goal is to transmit the information bits as far as possible from the source in the direction of destination and thus, the performance is best described by a measure of the number of information bits and how far they are communicated from the source in the direction of destination, both of which are characterized by the transmission rate and progress respectively.

Another key feature of a wireless network is spatial reuse, the ability to maintain simultaneous transmissions over different spatial regions of the network. Spatial reuse is characterized by the density of transmissions in the network. Based on the above discussed factors, the performance metric we use in the paper is \textit{Progress Rate Density} defined as \textit{the product of the number of information bits in bps communicated reliably per unit area of the network and the associated progress}. The PRD metric was introduced in \cite{Blomer_II} and an earlier version of PRD focused only on progress and density appeared in \cite{Baccelli}.
\section{System Model}
\label{sysmod}
We consider a wireless adhoc multihop network in which nodes are modeled as a 2-D homogeneous Poisson point process (PPP) $\Phi=\{i, X_i\}$ of intensity $\lambda$ $m^{-2}$, where $X_i$ denotes the coordinates of node $i$.
The MAC layer uses the spatial reuse ALOHA protocol\cite{Baccelli}.
The physical communication resource consists of orthogonal discrete time slots. In every time slot, a node $i \in \Phi$ either acts as a transmit node with medium access probability (MAP) $p$ or as a receive node with probability $1-p$. The decision process to be either a transmit or receive node is independent from slot to slot. A node $i \in \Phi$ makes transmit or receive decisions independent of other nodes in the network. The parent PPP $\Phi$ can be split into 2 independent PPP's $\Phi^t$ and $\Phi^r$ of intensities $\lambda p$ $m^{-2}$ and $\lambda(1-p)$ $m^{-2}$ respectively.

Each slot duration is split into two non-overlapping phases,
\begin{itemize}
\item In the $1^{st}$ phase, a node $\in \Phi^t$ transmits either its own data packet or a data packet of another source node. As per the terminology of section \ref{Multihop_relay}, a node $\in \Phi^t$ will either be a source node or a forwarding relay node.
\item In the $2^{nd}$ phase, all the nodes of $\Phi^r$ that decode the data packet based on the transmission in the $1^{st}$ phase \footnote{More precisely the current block of the codeword from the $1^{st}$ phase is combined with the previous blocks of the codeword received during previous slots as per incremental redundancy combining of section \ref{IR_com}.} participate in the distributed contention scheme of section \ref{Re_Se} to select the forwarding relay for next hop communication. All nodes of $\Phi^r$ are potential relay nodes.
\end{itemize}

We assume IID block fading across slots.
It is beyond the scope of this paper to give a detailed traffic model description. From \cite{Baccelli}, we use the basic traffic model assumptions of a mean value of $\tau$ fresh packets per slot per source-destination pair and a service rate of $p$ per node.

Due to the homogeneous PPP assumption, the performance of the entire network can be quantified by studying a reference source destination communication. Without loss of generality, we assume that node $0$ is the reference source. For simplicity, we consider the reference source to be located at the origin i.e., $X_0=(0,0)$. Node $n_d$ is the reference destination, where $n_d$ is a large positive integer. It is located at an asymptotic distance along the X-axis i.e., $X_{n_d}$ is a point on the positive X-axis at a large distance from the origin. The reference source destination pair is depicted in Fig.\ref{Ntw_snapshot2015}. Conditioning on the source node at the origin does not affect the distribution of the homogeneous PPP $\Phi$ (See Slivnyak's theorem \cite{Bacc_book1} for more details).

In the next subsection, we present an analytical framework for studying the cooperative relaying protocol employing incremental redundancy combining. For the ease of presentation, in the following we assume that each relay node in the cooperative relaying protocol is allowed to combine two blocks of the codeword for decoding. The extension to the case where a relay node combines $M(>2)$ blocks of the codeword is addressed later in the paper.

\subsection{Incremental Redundancy Combining (IRC)}
\label{ir_sub}
Source node at origin encodes an information packet of length $b$ bits into a $N$-symbol codeword. The codeword is split into two non-overlapping blocks of length $L=\frac{N}{2}$ by a puncturing process. The source transmits the $1^{st}$ block of the codeword at code rate $R=\frac{b}{L}$. An important property of the puncturing process to note is that the $1^{st}$ block of the codeword is sufficient to decode the information bit vector.

The received signal at a node $v \in \mathbb{R}^2$ based on the transmission from the source node at origin is given by
\begin{equation}
\mathrm{y}=h_{0}|v|^{-\alpha/2}\mathrm{x}_0+\sum_{k\in\Phi^t}h_{k}|v-X_k|^{-\alpha/2}\mathrm{x}_k+\mathrm{z}\label{Rx_eq}
\end{equation}
where $h_{k}\sim \mathcal{CN}\left(0,1\right)$ is the Rayleigh fading coefficient from transmit node $k$, $\mathrm{x}_k$ is the message symbol of transmit node $k$ and $\alpha$ is the path loss exponent. In (\ref{Rx_eq}), the $1^{st}$ term represents the desired signal, the $2^{nd}$ term represents the interference and $\mathrm{z}$ is the additive Gaussian noise.
The instantaneous SINR at receive node $v \in \mathbb{R}^2$ from the source node at origin is given by
\begin{equation}
\mathrm{SINR}\left(v,0\right)=\frac{\rho\abs{h_{0}}^2|v|^{-\alpha}}{\sum_{k\in\Phi^t}\rho\abs{h_{k}}^2|v-X_k|^{-\alpha}+\sigma^2},\label{SIR_eq}
\end{equation}
where $\sigma^2$ is the noise power and $\rho$ is the transmit power. In this paper, we focus on a multihop random network which has a large number of nodes. The network density will be in the interference limited regime where the effect of noise is negligible. Hence in the following we assume $\sigma^2=0$ and the quantity in (\ref{SIR_eq}) becomes $\mathrm{SIR}\left(v,0\right)$.

\begin{figure}[!hbtp]
\centering
\includegraphics[width=0.5\textwidth]{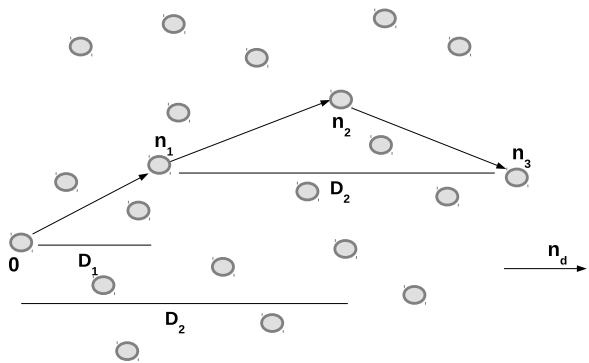}
\caption{Source (node $0$) transmits the $1^{st}$ block of the codeword. Node $n_1$ decodes the data packet using the $1^{st}$ block from node $0$ and offers the most progress $D_1$ from node $0$ in the direction of destination (node $n_d$), which is at an asymptotic distance from node $0$ along the x-axis. Node $n_2$ decodes the data packet by combining two blocks of the codeword received from nodes $\{0,n_1\}$. The node $n_2$ offers the progress $D_2$ from the source in the direction of the destination. Similarly the node $n_3$ combines two blocks of the codeword received from nodes $\{n_1,n_2\}$. (Without any ambiguity, we use the convention that node $n_i$ acts as $i^{th}$ relay.)}
\label{Ntw_snapshot2015}
\end{figure}

All the nodes $\in \Phi^r$ receive the $1^{st}$ block of the codeword from the source and make an attempt to decode the data packet. The relay nodes that successfully decode the data packet participate in a distributed contention scheme. From the definition of progress in section \ref{Re_Se} as the distance from the source in the source-destination direction over which the information bits are communicated, the progress of a relay in the above presented system model 
is the distance from origin along the positive X-axis over which the info bits are communicated. The distributed contention scheme selects the relay that offers the most progress as the forwarding relay. Let the node $n_1\in \Phi^r$ with coordinate $X_{n_1}$ be the forwarding relay. The relay selection is illustrated in Fig.\ref{Ntw_snapshot2015} which shows the reference source destination communication route when each relay node is allowed to combine two blocks of the codeword for decoding. The node $n_1$ offers the most progress from the origin from among the relay nodes which decode the data packet using $1^{st}$ block of the codeword from node $0$. Mathematically, the progress offered by the node $n_1$ is given by
\begin{equation}
\mathrm{D}_1= \max_{i\in\Phi^r}\Big[\mathbf{1}\big(I_1\left(X_i\right)\geq R\big)~\abs{X_i}\cos\left(\theta\left(X_i\right)\right)\Big]
\label{hop1}
\end{equation}
where $I_1\left(X_i\right)=\log_2\left(1+\mathrm{SIR}\left(X_i,0\right)\right)$ is the mutual information (MI) achieved by relay node $i$ based on the $1^{st}$ block of the codeword from node $0$, $\mathbf{1}\left(\cdot\right)$ is the indicator function and $\theta\left(\cdot\right)$ is the angle relative to positive $x-$axis. As mentioned earlier, the destination node is at an asymptotic distance along the X-axis and the expression for $1^{st}$ hop progress in (\ref{hop1}) considers the progress offered by each relay node along the X-axis direction as measured by the $\abs{X_i}\cos\left(\theta\left(X_i\right)\right)$ term.

Since the node $n_1$ was able to decode the data packet, it will regenerate the complementary block i.e., the $2^{nd}$ block of the codeword and transmit it in a future slot. In this paper, since the key focus is to measure how far the information bits are communicated from the source in the source-destination direction, we just assume that the forwarding relays transmit the blocks of the codeword within a few slots after they are selected.

During the $2^{nd}$ hop communication, the node $n_1$ transmits the $2^{nd}$ block of the codeword at rate $R$ in the $1^{st}$ phase of the slot it chooses to transmit. In the $2^{nd}$ phase of that slot, all the nodes $\in \Phi^r$  combine the $2^{nd}$ block of the codeword from node $n_1$ with the $1^{st}$ block of the codeword from the source \footnote{Some nodes $\in \Phi^r$ which have the $2^{nd}$ block of the codeword from node $n_1$ may not have the $1^{st}$ block from the source because they were not in receive mode when node $0$ was transmitting. In this case, these nodes use only $1$ block of the codeword for decoding.} and make an attempt to decode the data packet. Out of the successful relay nodes, the one with the most progress from origin is selected as the forwarding relay 2. The node $n_2\in \Phi^r$ with coordinate $X_{n_2}$ denotes the forwarding relay 2. The node $n_2$ is depicted in Fig.\ref{Ntw_snapshot2015} where it combines the $2$ blocks of the codeword from nodes $n_1$ and $0$ to decode the data packet and offers the most progress from origin along the positive X-axis.
Mathematically, the progress from the origin up to the node $n_2$ is given by
\begin{align}
\mathrm{D}_2&= \max_{i\in\Phi^r}\Bigg[\mathbf{1}\big(I_1\left(X_i\right) +I_2\left(X_i\right)\geq R\big)~
\abs{X_i}\cos\left(\theta\left(X_i\right)\right)\Bigg]
\label{hop2_t}
\end{align}
where $I_2\left(X_i\right)=\log_2\left(1+\mathrm{SIR} \left(X_i,X_{n_1}\right)\right)$ is the MI achieved by relay node $i$ based on the $2^{nd}$ block of the codeword from node $n_1$ and $I_1\left(X_i\right)$ is the MI based on the $1^{st}$ block of the codeword from node $0$.

The cooperative relaying continues with the node $n_3$ which decodes the
data packet by combining two blocks of the codeword received from nodes $n_1$ and $n_2$ and offers the most progress. This process continues until the data packet reaches the destination node $n_d$ and an ACK is sent back after successful decoding.

As mentioned in section \ref{per_met}, the performance metric used in this paper to study the multihop random network is progress rate density.
All forwarding relays transmit one block of the codeword at the same code rate $R$. The density of transmissions in the network is $\lambda p$ $m^{-2}$. The progress terms defined in (\ref{hop1}) and (\ref{hop2_t}) are random variables and hence we define an expected measure of the same to use in the performance metric.
The expected progress can be defined as
\begin{align}
d_2\left(R,p\right)&=\mathbb{E}\left[\mathrm{D}_2\right]
\label{aver_prg}\\
d_1\left(R,p\right)&=\mathbb{E}
  \left[\mathrm{D}_1\right]\label{Expe_D1}
\end{align}
where the $\mathbb{E}\left[\cdot\right]$ is taken w.r.t PPP $\Phi$.

As mentioned in section \ref{new_intro}, the performance of cooperative relaying protocol is compared to that of conventional relaying with no cooperation. Hence in the following, the performance metrics for relaying protocols with and without cooperation are defined.

\subsubsection{No Cooperation (NC)}
For a conventional relaying protocol with no cooperation, the progress rate density of the network is given by
\begin{equation}
\mathrm{PRD}= R~\lambda p~d_1\left(R,p\right)\label{C1_exp}
\end{equation}
\subsubsection{Cooperative Relaying}
$d_2$ in (\ref{C2_exp}) is a measure of progress which spans two hops. To compare the PRD of cooperative relaying protocol to (\ref{C1_exp}), we need a measure of progress per hop. So we define $d_2-d_1$ as the progress per hop by combining two blocks of the codeword at relay nodes. Hence for the cooperative relaying protocol with incremental redundancy combining, the progress rate density of the network is given by
\begin{equation}
\mathrm{PRD}= R~\lambda p~\big(d_2\left(R,p\right)- d_1\left(R,p\right)\big)\label{C2_exp}
\end{equation}

The cooperative relaying protocol presented in section \ref{Multihop_relay} utilizes incremental redundancy combining. In the following, we extend the cooperative relaying protocol to the case where the network employs a form of combining known as repetition combining in which an entire codeword is transmitted from the source node and repeated by the forwarding relay nodes. For reception, the relay nodes perform maximal ratio combining of the repeated codewords from current and previous transmissions extracting space and time diversity.

\subsection{Repetition Combining (RC)}
\label{rtd_sub}
The source node at origin encodes $b$ information bits into a $N$-symbol codeword. Both the source and the intermediate forwarding relays transmit the entire $N$-symbol codeword at the code rate $R=\frac{b}{N}$.

During the $1^{st}$ hop communication from source node to forwarding relay $1$ i.e., node $n_1$, the progress achieved is $\mathrm{D}_1$ given in (\ref{hop1}). During the $2^{nd}$ hop communication, the forwarding relay $1$ transmits the same $N$-symbol codeword as the source at rate $R$. The forwarding relay $2$ i.e., node $n_2$ maximal ratio combines the repeated codewords from nodes $\{0, n_1\}$ for decoding the data packet and offers the most progress from origin along the positive X-axis. The progress from origin upto the node $n_2$ is given by
\begin{align}
\mathrm{D}_2&= \max_{i\in\Phi^r}\Bigg[\mathbf{1}\bigg(\mathrm{SIR}\left(X_i,0\right)+
\mathrm{SIR}\left(X_i,X_{n_1}\right)\geq 2^R-1\bigg)\nonumber\\
&~~~\abs{X_i}\cos\left(\theta\left(X_i\right)\right)\Bigg]
\label{hop2_rtd}
\end{align}

The performance of the cooperative relaying protocol as described in the above section is measured by the PRD metric. To understand the operation of the cooperative relaying protocol at the maximal PRD point, the parameters of the protocol need to be optimized. The parameters that can be tuned are coding rate $R$ and MAP $p$.

The PRD in (\ref{C1_exp}) and (\ref{C2_exp}) is evaluated based on Monte Carlo simulation. In order to tune the cooperative relaying protocol based on optimization of analytic functions, we develop an analytic approximation to the PRD of the network in the next section. With a closed form expression for the PRD, the parameters of the protocol can be optimized numerically by optimization methods.
\section{PRD Approximation}
\label{appro_prd}
In this section, we develop an analytic approximation to the expected progress defined in (\ref{aver_prg}) and thus obtain an approximation to the PRD in (\ref{C2_exp}). It is conceptually infeasible to evaluate the distribution and expectation of the progress $D_2$ defined in (\ref{hop2_t}). Alternatively we develop a heuristic approximation to the expected progress $d_2$.
The approximation developed is based on the concept of decoding cells introduced in \cite{Bacc_book1}. Decoding cells in their simplest form are areas in $\mathbb{R}^2$ containing points with successful reception of data packets from the origin and are more thoroughly defined in the following.
We first develop the approximation to the expected progress for incremental redundancy combining and then present the same for repetition combining.

\subsection{Incremental Redundancy}
\label{ir_Capa_Opt}
In the following, the decoding cell for incremental redundancy combining is formally defined.
\subsubsection{Decoding Cell}
\label{C_2def}
A decoding cell $\mathrm{\Sigma}_2$ is defined as
\begin{align}
&\mathrm{\Sigma}_2= \left\lbrace v \in \mathbb{R}^2:~I_1+I_2 \geq R \right\rbrace\nonumber\\
&I_1=\log_2\left(1+\mathrm{SIR}\left(v,0\right)\right),\nonumber\\
&I_2=\log_2\left(1+\mathrm{SIR}\left(v,\eta_1\right)\right),\nonumber\\
&\eta_1=\left(\tilde{d}_1,0\right)\label{cell_def}
\end{align}
where $\tilde{d}_1$ is an approximation to the expected progress $d_1$ in (\ref{Expe_D1}). $\tilde{d}_1$ has a closed form expression as a function of system parameters but for the ease of presentation, the expression is presented later.

The cell $\mathrm{\Sigma}_2$ contains all $v \in \mathbb{R}^2$ that decode the data packet using two blocks of the codeword from origin and $\eta_1$ respectively. The point $\eta_1$ in (\ref{cell_def}) represents the equivalent of the location of forwarding relay $1$. Although the progress in (\ref{hop2_t}) involves the instantaneous random location of forwarding relay $1$, in the definition of cell $\mathrm{\Sigma}_2$ we use an approximate location given by $\eta_1$ for analytical tractability. The coordinate-1 of forwarding relay $1$ is given in (\ref{hop1}). Since we are interested in the expected location of the forwarding relay for cell definition, we set the coordinate-1 of $\eta_1$ to $\tilde{d}_1$. Now there is no information about the coordinate-2 of forwarding relay $1$. Also we are only interested in the progress from origin along the positive $X$-axis. Hence using PPP stationarity to simplify the analysis, we set the coordinate-2 of $\eta_1$ to $0$. Such a point $\eta_1$ will be useful to compute a tractable and valuable approximation to the expected progress $d_2$ in (\ref{aver_prg}) and will be explained shortly.

The average cell area is given by
\begin{equation}
\mathbb{E}\big[\abs{\mathrm{\Sigma}_{2}}\big]=\int_{\mathbb{R}^2}
\mathbb{P} \left(I_1+I_2 \geq R\right)~\ud v\label{are_eq}
\end{equation}
An interpretation of the average cell area is that it contains all $v \in \mathbb{R}^2$ which in the expected sense can decode the data packet using two blocks of the codeword. By homogeneity of the PPP $\Phi$, the relay nodes in the average cell area are uniformly distributed. Using these properties, the following theorem derives an approximate expression for the expected progress $d_2\left(R,p\right)$ in (\ref{aver_prg}).

\begin{figure}[!hbtp]
\centering
\includegraphics[width=0.5\textwidth]{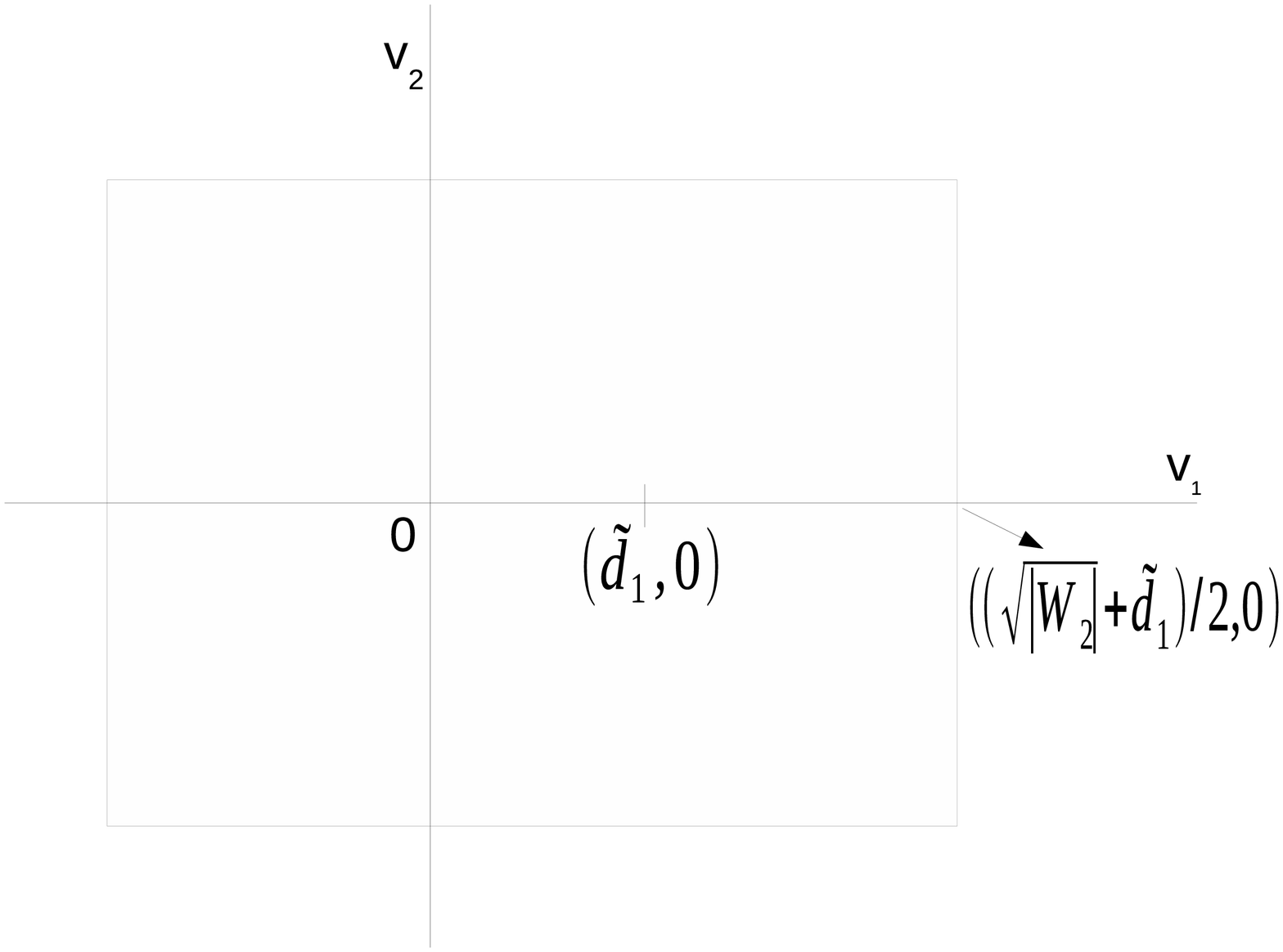}
\caption{A square $W_2$ centered around the two points origin and $\eta_1=\left(\tilde{d}_1,0\right)$ represents the decoding cell $\mathrm{\Sigma}_2$. The center of the square is $\left(\tilde{d}_1/2,0\right)$. The maximum progress offered by nodes of $\Phi^r$ in $W_2^+$, the portion of $W_2$ in the positive $v_1$ axis, is $\left(\sqrt{\abs{W_2}}+\tilde{d}_1\right)/2$.}
\label{Cell_Diag}
\end{figure}
\begin{Theoi1}
\label{The_1}
An approximation to the expected progress of cooperative relaying protocol with incremental redundancy combining $\tilde{d}_2\left(R,p\right)$ is given by
\begin{IEEEeqnarray}{rCl}
\tilde{d}_2\left(R,p\right)&=& \frac{\sqrt{\abs{W_2}}+\tilde{d}_1}{2}~ \left(1-\frac{1-e^{-c_2}}{c_2}\right)\label{d2_appr}\\
c_2&=&\frac{\lambda (1-p)}{2}\left(\abs{W_2}+\tilde{d}_1 \sqrt{\abs{W_2}}\right)\nonumber\\
\abs{W_2}&\geq&\frac{\pi}{A} \Bigg[\frac{2}{T}-\frac{\exp\left(
\frac{-AT \tilde{d}_1^2}{1+T}\right)}{1+T}+\frac{\exp\left(
\frac{-A\bar{T} \tilde{d}_1^2}{1+\bar{T}}\right)}{1+\bar{T}} -\frac{\exp\left(\frac{-AT\bar{T} \tilde{d}_1^2} {T+\bar{T}}\right)} {T+\bar{T}}\Bigg]\label{w2_lb}
\end{IEEEeqnarray}
where $T=\left(2^R-1\right)^{\delta}$, $\bar{T}=\left(2^R/2-1\right)^{\delta}$, $\delta=2/\alpha$ and $A=\lambda p G\left(\alpha\right)$, $G\left(\alpha\right)=\tfrac{\pi\delta}{\sin\left(\pi\delta\right)}$.
\end{Theoi1}
$\tilde{d}_1$ is an approximation to expected progress $d_1$. A closed form expression for $\tilde{d}_1$ is derived based on a decoding cell defined with only origin as the center.
From \cite{Blomer_II}, the approximation $\tilde{d}_1$ is given by
\begin{align}
&\tilde{d}_1=\frac{\sqrt{\abs{W_1}}}{2}~\left(1-\frac{1-e^{-c_1}}{c_1}
\right)\label{d1_exp}\\
&~~c_1=\lambda (1-p)\abs{W_{1}}/2,~~\abs{W_{1}}=\frac{\pi}{\lambda
p G\left(\alpha\right)T}\nonumber
\end{align}
\begin{IEEEproof}
Consider a square $W_2$ centered around the two points origin and $\eta_1$ with area $\abs{W_2}=\mathbb{E}\big[\abs{\mathrm{\Sigma}_2}\big]$ as shown in Fig.\ref{Cell_Diag}.
Let $W_2^+$ represent the portion of $W_2$ in the positive $v_1$ axis. The area of $W_2^+$ is given by $\abs{W_2^+}=\left(\abs{W_2}+\tilde{d}_1 \sqrt{\abs{W_2}}\right)\big/2$.

Define $G$ as the number of nodes of PPP $\Phi^r$ in $W_2^+$. By PPP stationarity, $G=\abs{\Phi^r\left(W_2^+\right)}$ is Poisson distributed with parameter $c_2=\lambda (1-p)\abs{W_2^+}$.
The nodes of $\Phi^r$ in $W_2^+$ offer a maximum progress of $\left(\sqrt{\abs{W_2}}+\tilde{d}_1\right)\big/2$.

Hence based on the above mentioned properties, an approximate expression for $\mathbb{E}\big[\mathrm{D}_2\big]$ is given by
\begin{align}
\mathbb{E}\big[\mathrm{D}_2\big]\approx \tilde{d}_2\left(R,p\right)&= \sum_{k=0}^{\infty}\mathbb{E}\Big[\max_{i\leq k}~U_{i,v_1}\big |
G=k\Big]~ \mathbb{P}\left(G=k\right)\label{int_eq_prg}\\
&=\sum_{k=0}^{\infty} \frac{\sqrt{\abs{W_2}}+\tilde{d}_1}{2}~ \frac{k}{k+1}~\mathbb{P}\left(G=k\right)\nonumber\\
&=\frac{\sqrt{\abs{W_2}}+\tilde{d}_1}{2}~\sum_{k=0}^{\infty}
\mathbb{P}\left(G=k\right)\left(1-\frac{1}{k+1}\right)\nonumber\\
&=\frac{\sqrt{\abs{W_2}}+\tilde{d}_1}{2}~\left(1-\frac{1-e^{-c_2}}{c_2}
\right)\label{cell_appr}
\end{align}
where in (\ref{int_eq_prg}) $U_{i,v_1}$ are uniformly distributed over $\left[0, \left( \sqrt{\abs{W_2}}+\tilde{d}_1\right)/2 \right]$ and (\ref{cell_appr}) follows from the previous line after plugging in the expression for $\mathbb{P}\left(G=k\right)$ in the $2^{nd}$ sum and
simplifying it further.

To complete the proof, we need to provide an expression for the square area $\abs{W_2}$, which is the same as the average cell area $\mathbb{E}\big[\abs{\mathrm{\Sigma}_{2}}\big]$ defined in (\ref{are_eq}).
We now focus on the derivation of $\mathbb{E}\big[\abs{\mathrm{\Sigma}_2}\big]$ expression. In Appendix \ref{der_exp_ir}, the lower bound for $\mathbb{E}\big[\abs{\mathrm{\Sigma}_2}\big]$ given in (\ref{w2_lb}) is derived.
\end{IEEEproof}

Based on the approximation to expected progress $d_2$, an approximate analytic expression for the PRD in (\ref{C2_exp}) is given by
\begin{equation}
\tilde{\mathrm{PRD}}=R~\lambda p~\Big[\tilde{d}_2\left(R,p\right)- \tilde{d}_1\left(R,p\right)\Big]\label{Cap_exp}
\end{equation}

\subsection{Repetition Combining}
\label{rtc_Opt}
In the following subsection, the analytic approximation to PRD for the case of repetition combining is developed. The approximation is developed by following the same steps as in section \ref{ir_Capa_Opt}. We start out by formally defining the decoding cell for repetition combining.
\subsubsection{Decoding Cell}
\label{Cell_def_rtc}
The decoding cell $\mathrm{\Sigma}_2$ is defined as
\begin{align}
\mathrm{\Sigma}_{2}= &\left\lbrace v \in \mathbb{R}^2: ~\mathrm{SIR}\left(v,0\right)+\mathrm{SIR}\left(v,\eta_1\right) \geq 2^R-1 \right\rbrace\label{cell_def_rtd}
\end{align}
The cell $\mathrm{\Sigma}_2$ contains all $v \in \mathbb{R}^2$ that decode the data packet by maximal ratio combining the repeated codewords from origin and $\eta_1$.
The average cell area is given by
\begin{equation}
\mathbb{E}\big[\abs{\mathrm{\Sigma}_2}\big]=\int_{\mathbb{R}^2}
\mathbb{P} \left(\mathrm{SIR}\left(v,0\right)+ \mathrm{SIR}\left(v,\eta_1\right) \geq 2^R-1\right)~\ud v\label{are_eq_rtd}
\end{equation}
Based on the average cell area in (\ref{are_eq_rtd}) and the properties of PPP $\Phi^r$, the following theorem provides an approximate expression for expected progress $d_2$.
\begin{Theoi2}
An approximation to the expected progress of cooperative relaying protocol with repetition combining $\tilde{d}_2\left(R,p\right)$ is given by
\begin{align}
\tilde{d}_2\left(R,p\right)&= \frac{\sqrt{\abs{W_2}}+\tilde{d}_1}{2}~ \left(1-\frac{1-e^{-c_2}}{c_2}\right)\label{d2_appr_rtc}\\
c_2&=\frac{\lambda (1-p)}{2}~ \left(\abs{W_2}+\tilde{d}_1\sqrt{\abs{W_2}}\right)\nonumber\\
\abs{W_2}&\geq\frac{\pi}{A} \Bigg[\frac{2}{T}-\frac{\exp\left(
\frac{-AT \tilde{d}_1^2}{1+T}\right)}{1+T}+\frac{\exp\left(
\frac{-A\tilde{T} \tilde{d}_1^2}{1+\tilde{T}}\right)}{1+\tilde{T}} -\frac{\exp\left(\frac{-AT\tilde{T} \tilde{d}_1^2} {T+\tilde{T}}\right)} {T+\tilde{T}}\Bigg]\label{w2_lb_rc}
\end{align}
where $\tilde{T}=\left(2^R-2\right)^{\delta}$.
\end{Theoi2}
\begin{IEEEproof}
The proof of theorem 2 follows the same ideas as in theorem \ref{The_1}.
Define a square $W_2$ of area equal to $\mathbb{E}\big[\abs{\mathrm{\Sigma}_2}\big]$ in (\ref{are_eq_rtd}). Using the stationary property of PPP $\Phi^r$ and following the same steps as in theorem \ref{The_1} leads to (\ref{d2_appr_rtc}).

The only difference between (\ref{d2_appr_rtc}) and (\ref{d2_appr}) is in the area of square $W_2$. For the repetition combining case, the average area of the decoding cell $\Sigma_2$ is discussed below.

The lower bound for $\mathbb{E}\big[\abs{\mathrm{\Sigma}_2}\big]$ given in (\ref{w2_lb_rc}) is derived in Appendix \ref{der_exp_rtd} and this completes the proof.
\end{IEEEproof}
\section{PRD Optimization}
\label{opti_prd}
The cooperative relaying protocol described in section \ref{Multihop_relay} transports the data packets from the source node to destination via multiple hops either using incremental redundancy combining or repetition combining at the intermediate relay nodes. The performance of the protocol is studied through the PRD metric. The PRD measures the following quantities: the amount of information communicated in bps, how far from the source in the source-destination direction this information is communicated and the spatial reuse factor i.e., the number of transmissions per unit area of the network.
For the system model in the paper, these three quantities are directly determined by the parameters of the cooperative relaying protocol such as coding rate $R$ and MAP $p$. Choosing a higher $R$ increases the amount of information communicated but sacrifices the progress of information bits from the source. A large progress of information bits can be achieved by choosing low $R$ at the expense of the amount of information communicated.
The MAP $p$ influences the spatial reuse factor $\lambda p$ and the progress of information bits from the source. Choosing a high $p$ increases the spatial reuse in the network whereas the increased interference leads to smaller progress of information bits. A similar tradeoff is observed for a low value of $p$.

Hence for the cooperative relaying protocol to function efficiently and to have performance gains over conventional relaying, the protocol needs to be tuned. The parameters $R$ and $p$ need to be optimized to operate the network at the maximal PRD point. Maximization of the PRD in (\ref{C2_exp}) and its analytic approximation in (\ref{Cap_exp}) is discussed in the following.
\subsection{Exact PRD Maximization}
The optimal MAP $p$ and coding rate $R$ are given by
\begin{equation}
    \langle R,p\rangle=\arg \max_{R,p}~~R~\lambda p~\big(d_2\left(R,p\right)- d_1\left(R,p\right)\big)\label{Opt_va}
\end{equation}
Both optimal $R$ and $p$ are solved by Monte Carlo simulation and the numerical results are presented in section \ref{sim_res}.

\subsection{Approximate PRD Maximization}
The coding rate $\tilde{R}$ and MAP $\tilde{p}$ that maximize the approximate PRD are given by
\begin{equation}
    \langle \tilde{R},\tilde{p}\rangle=\arg \max_{R,p}~~ R~\lambda p~\Big[\tilde{d}_2\left(R,p\right)- \tilde{d}_1\left(R,p\right)\Big]\label{Apr_va}
\end{equation}
The objective function is concave and the KKT points are solved by gradient descent methods.

\section{Cooperative Relaying with $M>2$}
\label{hig_CooRe}
The cooperative relaying protocol can be extended to the case where each relay node combines $M$ blocks of a codeword to decode the data packet. The $M$ blocks of the codeword are received from the current forwarding relay and the previous $M-1$ forwarding relays. As mentioned in section \ref{IR_com}, the source and forwarding relays $1, 2, \cdots M-1$ transmit the first $M$ blocks of the codeword. Applying this to the system model presented in this paper, the source and nodes $\{n_1, n_2, \cdots n_{M-1}\}$ transmit the first $M$ blocks of the codeword respectively. From forwarding relay $M$ i.e., node $n_M$ onwards, the transmitted block of the codeword is complementary to the $M-1$ most recent blocks of the codeword received. More precisely, the node $n_i$ transmits the $\left(q(i)+1\right)^{th}$ block of the codeword where $q(i)=\mod\left(i,M\right)$.

The progress from origin up to the node $n_M$ is given by
\begin{equation}
\mathrm{D}_M= \max_{i\in\Phi^r}\Bigg[\mathbf{1}\Big(\sum_{k=1}^M I_k\left(X_i\right)\geq R\Big)~\abs{X_i}\cos\left(\theta\left(X_i\right)\right)\Bigg]
\label{prg_M}
\end{equation}
where $I_k\left(X_i\right)=\log_2\left(1+\mathrm{SIR}\left(X_i,X_{n_{k-1}}\right)
\right)$ is the MI achieved by relay node $i$ based on the $k^{th}$ block of the codeword from node $n_{k-1}$. The expected progress is given by $d_M\left(R,p\right)=\mathbb{E}
\left[\mathrm{D}_M\right]$.

The PRD of the cooperative relaying protocol where each relay node combines $M$ blocks of a codeword is given by
\begin{equation}
\mathrm{PRD}=R~\lambda p~\big(d_M\left(R,p\right)-d_{M-1}\left(R,p\right)\big)\label{PRD_expM}
\end{equation}
\subsection{Decoding Cell}
\label{prd_Mdef}
The decoding cell for the cooperative relaying protocol where each relay node combines $M$ blocks of a codeword $\mathrm{\Sigma}_M$ is defined as
\begin{IEEEeqnarray}{rCl}
\mathrm{\Sigma}_M&=&\left\lbrace v \in \mathbb{R}^2:~\sum_{k=1}^M I_k \geq R \right\rbrace\label{cell_defM}\\
I_k&=&\log_2\left(1+\mathrm{SIR}\left(v,\eta_{k-1}\right)\right),\nonumber\\
\eta_{k-1}&=&\left(\tilde{d}_{k-1},0\right)\label{Ek1}
\end{IEEEeqnarray}
The average area of the decoding cell $\mathrm{\Sigma}_M$ is given by
\begin{equation}
\mathbb{E}\big[\abs{\mathrm{\Sigma}_{M}}\big]=\int_{\mathbb{R}^2}
\mathbb{P} \Big(\sum_{k=1}^M I_k \geq R\Big)~\ud v \label{avg_areaM}
\end{equation}

\subsection{Approximation to Expected Progress}
\label{appr_prgM}
Similar to Theorem 1, we define a square $W_M$ of area equal to $\mathbb{E}\big[\abs{\mathrm{\Sigma}_{M}}\big]$ centered around two points origin and $\eta_{M-1}$. The nodes of PPP $\Phi^r$ in the portion of $W_M$ to the positive $v_1$ axis offer a maximum progress of $\left(\sqrt{\abs{W_M}}+\tilde{d}_{M-1}\right)\big/2$. Using the property that the number of nodes of PPP $\Phi^r$ in the portion of $W_M$ to the positive $v_1$ axis is Poisson distributed, the approximation to expected progress $d_M$ is given by
\begin{IEEEeqnarray}{rCl}
\tilde{d}_M\left(R,p\right)&=& \frac{\sqrt{\abs{W_M}}+\tilde{d}_{M-1}}{2}~ \left(1-\frac{1-e^{-c_M}}{c_M}\right)\label{dM_appr}\\
c_M&=&\frac{\lambda (1-p)}{2}\left(\abs{W_M}+\tilde{d}_{M-1} \sqrt{\abs{W_M}}\right)
\end{IEEEeqnarray}
Based on (\ref{dM_appr}), the approximation to PRD in (\ref{PRD_expM}) can be defined accordingly.
\subsection{Repetition Combining}
\label{RC_sec}
A cooperative relaying protocol employing repetition combining with $M>2$ can be defined in a similar manner. Every relay node combines the $M$ transmissions from current and previous forwarding relays for decoding a packet. The progress term in (\ref{prg_M}) and the decoding cell in (\ref{cell_defM}) can be used to study the cooperative relaying protocol with repetition combining by replacing the MI sum with the corresponding sum of SIR's. The PRD approximation for the repetition combining case is defined based on (\ref{dM_appr}).
\section{Numerical Results}
\label{sim_res}
In this section, we present numerical results illustrating the performance of the cooperative relaying protocol proposed in the paper.
A wireless adhoc multihop network where nodes are distributed according to a 2-D homogeneous PPP of intensity
$\lambda$ $m^{-2}$ was simulated\cite{book1,Haenggi}. We assume that nodes experience Rayleigh fading, which is IID across slots and nodes. The network performance is measured by simulating the reference source destination communication. The following values of network parameters were used in the simulation, the network density $\lambda=1$ and the path loss exponent $\alpha=[2.5,4]$.

\begin{figure}[!hbtp]
\centering
\includegraphics[width=0.5\textwidth]{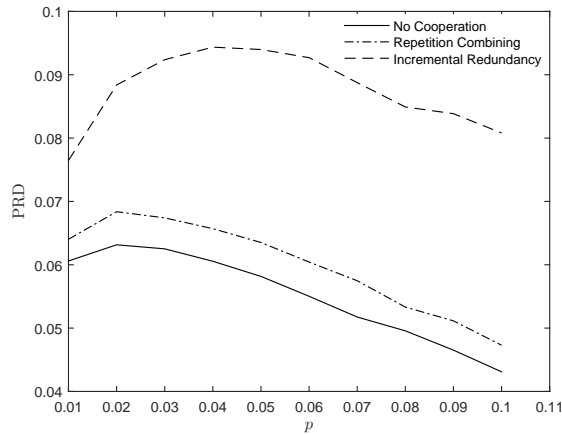}
\caption{Progress Rate Density $\mathrm{PRD}$ plotted as a function of MAP $p$ for relaying protocols with no cooperation, repetition combining and incremental redundancy combining respectively at $\lambda=1$, $R=3$ and $\alpha=3$.}
\label{Coop_Relay_benefit}
\end{figure}

Fig.\ref{Coop_Relay_benefit} shows a plot of the progress rate density PRD as a function of the MAP $p$ for $R=3$ at $\lambda=1$ and $\alpha=3$. The performance curve of a relaying protocol with no cooperation is plotted based on (\ref{C1_exp}). The curves for both incremental redundancy combining and repetition combining with diversity order $M=2$ are plotted from (\ref{C2_exp}). In conventional relaying with no cooperation, each relay node has access only to the transmission from the current forwarding relay. The relay nodes have to decode the information packet based on only this one transmission. On the other hand, in the case of both incremental redundancy combining and repetition combining with $M=2$, the relay nodes will have access to the transmission from the current forwarding relay and also the transmission from the previous forwarding relay. The relay nodes combine the two transmissions for decoding, thus extracting the space and time diversity inherent in the network leading to a higher throughput compared to the relaying protocol with no cooperation.

In incremental redundancy combining, every forwarding relay supplies new parity symbols to decode the information packet. These new parity symbols in addition to the available space time diversity enable the relay nodes to decode more information bits per packet and thus achieve a higher network throughput (PRD) compared to repetition combining. This effect is illustrated in the performance curves of Fig.\ref{Coop_Relay_benefit}. For example, when using conventional relaying with no cooperation the network achieves a PRD of $0.06055$ at $p=0.04$. Adding cooperative relaying in the form of repetition combining increases the PRD by $9\%$ whereas incremental redundancy combining leads to a $56\%$ increase in the network PRD.
\subsection{Maximization of PRD}
\label{prd_max}
\begin{figure}[!hbtp]
\centering
\includegraphics[width=0.5\textwidth]{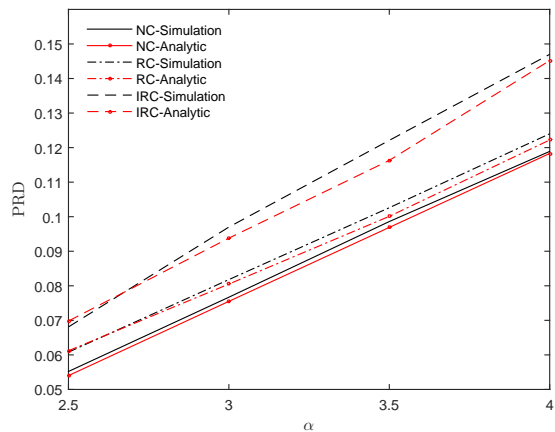}
\caption{A plot of progress rate density PRD values against the path loss exponent $\alpha$ for relaying protocols with no cooperation, repetition combining and incremental redundancy combining at $\lambda=1$. The PRD is maximized as per (\ref{Opt_va}) and (\ref{Apr_va}).}
\label{Coop_Relay_PRD_alpha}
\end{figure}

We now present the numerical results of the maximization of PRD as per (\ref{Opt_va}) and (\ref{Apr_va}). The PRD maximization in (\ref{Opt_va}) is solved by monte carlo simulations. Fig.\ref{Coop_Relay_PRD_alpha} shows a plot of the maximized progress rate density (PRD) values against the path loss exponent $\alpha$ for relaying protocols with no cooperation, repetition combining and incremental redundancy combining at $\lambda=1$. Cooperative relaying in the form of incremental redundancy combining leads to a near constant gain in network throughput at varying values of $\alpha$. From the performance curve for incremental redundancy combining, at $\alpha=3$ the network has a $26.5\%$ gain in PRD and at $\alpha=4$, the gain is $23.5\%$. As $\alpha$ decreases, the effect of interference in the network increases and thus the benefit of doing cooperative relaying increases, although the change is nominal. This nature of variation of the PRD gain as a function of $\alpha$ is also valid for repetition combining. For both conventional relaying with no cooperation and the two cooperative relaying techniques, the results from the simulation based optimization in (\ref{Opt_va}) and the analytic function based optimization in (\ref{Apr_va}) have a very close match.
\subsection{Effect of Diversity Order $M$}
\begin{figure}[!hbtp]
\centering
\includegraphics[width=0.5\textwidth]{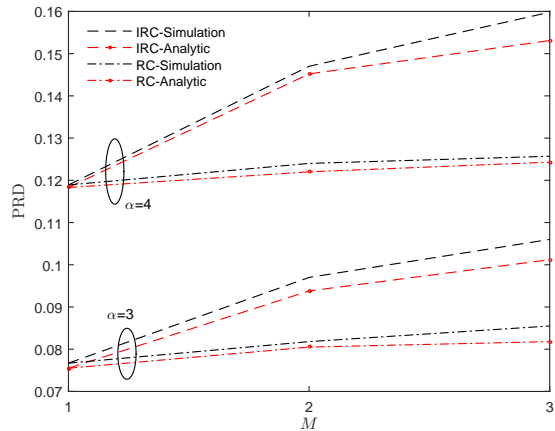}
\caption{The progress rate density of the network (PRD) plotted as a function of the diversity order $M$ for both repetition combining and incremental redundancy combining at $\lambda=1$ and $\alpha=\{3,4\}$. The PRD values are maximized based on (\ref{Opt_va}) and (\ref{Apr_va}).}
\label{Coop_Relay_PRD_M}
\end{figure}

In this subsection, we study how the performance of the cooperative relaying protocol varies with the diversity order $M$. Fig.\ref{Coop_Relay_PRD_M} shows a plot of the network PRD as a function of the diversity order $M$ for both repetition combining and incremental redundancy combining at $\lambda=1$ and $\alpha=\{3,4\}$. The PRD values are maximized based on (\ref{Opt_va}) and (\ref{Apr_va}). For cooperative relaying with $M>2$, the signal strengths i.e., SIR of the different transmissions that a relay node combines to decode a packet are non identical. For example, consider the typical source destination communication when $M=3$. The relay node $n_3$ combines three transmissions from the forwarding relay nodes $\{n_2, n_1, 0\}$ which are of decreasing strength on average due to the increasing distance from $n_3$. As a result of this decreasing signal strength of the transmissions, the benefit of cooperative relaying in terms of PRD gain becomes monotonic with the diversity order $M$. From the performance curve for incremental redundancy combining in Fig.\ref{Coop_Relay_PRD_M}, it is observed that at $\alpha=3$ the PRD increases by $26.5\%$ when the diversity order changes from $M=1$ to $M=2$ but when $M$ goes from $M=2$ to $M=3$, the PRD gain is only $9.3\%$. Such a monotonic nature of increase of the PRD with $M$ is consistent at $\alpha=3$ and also holds for repetition combining.

\subsection{Benefit of Analytic approximation}
To operate the network efficiently, the parameters of the cooperative relaying protocol need to be optimized. The protocol parameters coding rate $R$ and MAP $p$ can be optimized based on Monte Carlo simulations as per (\ref{Opt_va}). But this approach requires extensive computing time and resources. For example, computing one point on the curves shown in Figs.\ref{Coop_Relay_PRD_alpha} and \ref{Coop_Relay_PRD_M} can take upto tens of hours of simulation time on a state of the art PC. On the other hand, the optimization in (\ref{Apr_va}) based on the analytic approximation to PRD developed in section \ref{appro_prd} helps to perform protocol optimization with virtually no computing time and resources. The analytic function optimization of (\ref{Apr_va}) can be solved on the order of few seconds.
The PRD values of the network when operated at the $R$ and $p$ given by (\ref{Opt_va}) and (\ref{Apr_va}) are shown in Figs.\ref{Coop_Relay_PRD_alpha} and \ref{Coop_Relay_PRD_M}. Based on the proximity of the curves in both figures, we observe that the analytic function based optimization in (\ref{Apr_va}) facilitates to operates the network at a point very close to the optimal PRD point obtained from (\ref{Opt_va}).


One additional advantage of the analytic approximation based optimization is that once the close to optimal $R$ and $p$ are obtained from (\ref{Apr_va}), a second order search can be performed by a simulation around this small extremal region to arrive at the desired $R$ and $p$ for network operation. This approach takes much less computing time and resources than the pure simulation based search in (\ref{Opt_va}).
Performance of the analytic optimization in (\ref{Apr_va}) can possibly be further improved by computing the area $W_M$ efficiently compared to the bounds given in (\ref{w2_lb}) and (\ref{w2_lb_rc}).
\section{Conclusion}
\label{con_cl}
In this paper, we propose a cooperative relaying protocol for a large multihop network. The cooperative relaying protocol can use two forms of cooperative diversity namely incremental redundancy combining and repetition combining. Nodes of the multihop network are modeled by a homogeneous PPP with Rayleigh fading, constant power transmission per node and use slotted ALOHA protocol with spatial reuse as the MAC protocol. The performance of the cooperative relaying protocol is quantified by an analytic approximation to the progress rate density of the network. The performance gain of cooperative relaying protocol over conventional relaying with no cooperation is characterized at representative scenarios of the system parameters. It was shown that incremental redundancy combining provides higher gains in network throughput compared to repetition combining. The PRD gain due to
cooperative relaying protocol is monotonic in diversity order $M$ and has a near constant value as a function of path loss exponent $\alpha$.

A key assumption in the present paper is that all users transmit with the same power. Future direction would be to consider the possibility of adapting the transmit power of every user to maximize the network performance.
The results of the present paper suggest that gains in network performance can be achieved through cooperative relaying. However cooperative relaying was studied in a network where relays use fixed rate coding to transmit information. In \cite{JourVer}, it is shown that rateless codes achieve a higher rate density in the network compared to fixed rate codes. Hence an interesting future direction would be to study and examine the performance of a cooperative relaying protocol from source to destination with rateless coding employed at the source and relay nodes.
\section{Incremental Redundancy Combining}
\label{der_exp_ir}
We first evaluate the probability $\mathbb{P} \left(\cdot\right)$ in (\ref{are_eq}) and then subsequently derive an expression for $\mathbb{E}\big[\abs{\mathrm{\Sigma}_2}\big]$.
\subsection{Expression for $\mathbb{P} \left(I_1+I_2 \geq R\right)$}
From (\ref{cell_def}),
\begin{align}
&I_1=\log_2\left(1+\mathrm{SIR}\left(v,0\right)\right)\triangleq \log_2\left(1+S\abs{v}^{-\alpha}\right),\nonumber\\
&S=\frac{\abs{h_{0}}^2}{\sum_{k\in\Phi^t}\abs{h_{k}}^2|v-X_k|^{-\alpha}}\label{SIR_not}
\end{align}
The RV $S$ has been defined to assist in the derivation and is illustrated below.
\begin{align}
&\mathbb{P} \left(I_1+I_2 \geq R\right)\nonumber\\
&=\mathbb{P}\left(\log_2\left(1+S_1\abs{v}^{-\alpha}\right)+\log_2\left(1+S_2|v-\eta_1 |^{-\alpha}\right)\geq R\right)\nonumber\\
&=\int_0^{\infty}\mathbb{P}\left(\log_2\left(1+S_2|v-\eta_1 |^{-\alpha}\right)\geq \log_2\left(\frac{2^R}{1+s_1\abs{v}^{-\alpha}}\right)\right)~f_{S_1}\left(s_1\right)\ud s_1\nonumber\\
&=\int_0^{\infty}\mathbb{P}\left(S_2\geq \abs{v-\eta_1}^{\alpha}\left(\frac{2^R}{1+s_1 \abs{v}^{-\alpha}}-1\right)\right)~f_{S_1}\left(s_1\right)\ud s_1
\end{align}
From \cite{Baccelli}, the CCDF of RV $S$ is given as
\begin{align}
\mathbb{P}\left(S \geq s\right)=\left\{
\begin{array}{rl}
e^{-\lambda p G\left(\alpha\right)s^{\delta}} & ,s \geq 0\\
1 & ,s < 0.
\end{array}
\right. \label{S_ccdf}
\end{align}
Using (\ref{S_ccdf}) and defining $A=\lambda p G\left(\alpha\right)$ and $T=\left(2^R-1\right)^{\delta}$, we get
\begin{align}
&\mathbb{P} \left(I_1+I_2 \geq R\right)\nonumber\\
&=\int_0^{T^{\frac{1}{\delta}}\abs{v}^{\alpha}}e^{-A\abs{v-\eta_1}^2 \left(\frac{2^R}{1+s_1\abs{v}^{-\alpha}}-1\right)^{\delta}} f_{S_1}\left(s_1\right)~\ud s_1 +\int_{T^{\frac{1}{\delta}}\abs{v}^{\alpha}}^{\infty}f_{S_1}\left(s_1\right)
\ud s_1\nonumber\\
&\overset{(a)}{=}\int_0^{T^{\frac{1}{\delta}}\abs{v}^{\alpha}} e^{-A\abs{v-\eta_1}^2\left(\frac{2^R} {1+s_1\abs{v}^{-\alpha}}-1\right)^{\delta}}f_{S_1}\left(s_1\right)
~\ud s_1+e^{-AT\abs{v}^2}\label{Pr_in_te}\\
&\triangleq P_1+e^{-AT\abs{v}^2}
\end{align}
where (a) follows by evaluating the tail probability of RV $S_1$ at $T^{\frac{1}{\delta}}\abs{v}^{\alpha}$. Using the fact that pdf $f_{S_1}\left(s_1\right)$ follows from (\ref{S_ccdf}), the integral term in (\ref{Pr_in_te}) is written as
\begin{align}
P_1&=\int_0^{T^{\frac{1}{\delta}}\abs{v}^{\alpha}}e^{-A\abs{v-\eta_1}^2\left(\frac{2^R}{1+s_1\abs{v}^{-\alpha}}-1\right)^{\delta}}e^{-As^{\delta}_1}~
A~\ud\left(s^{\delta}_1\right)\nonumber\\
&\overset{(b)}{=}A\abs{v}^2\int_0^{T}e^{-A\abs{v-\eta_1}^2\left(\frac{2^R}  {1+\tau^{1/\delta}}- 1\right)^{\delta}} e^{-A\abs{v}^2\tau}~\ud \tau \label{P_1equ}
\end{align}
where (b) follows by the substitution $\tau=\abs{v}^{-2}s^{\delta}_1$.

Let $P_1\triangleq P_{1,a}+P_{1,b}$ with $P_{1,a}$ and $P_{1,b}$ defined as
\begin{align}
P_{1,a}&=A\abs{v}^2\int_0^1 e^{-A\abs{v-\eta_1}^2\left(\frac{2^R}  {1+\tau^{1/\delta}}- 1\right)^{\delta}} e^{-A\abs{v}^2\tau}~\ud \tau \nonumber\\
&\overset{(c)}{\geq} A\abs{v}^2\int_0^1 e^{-A\abs{v-\eta_1}^2\left(2^R-1\right)^{\delta}} e^{-A\abs{v}^2\tau}~\ud\tau\nonumber\\
&\overset{(d)}{=}e^{-A\abs{v-\eta_1}^2T} \left(1-e^{-A\abs{v}^2}\right)\label{I_pr}
\end{align}
where (c) is based on the fact that $\left(\frac{2^R}{1+\tau^{1/\delta}}- 1\right)^{\delta}$ is decreasing in $0\leq \tau < 1$ and (d) follows by taking the CDF of an exponential RV with parameter $A\abs{v}^2$ at $1$.
Similar bounds yield a lower bound for $P_{1,b}$.
\begin{align}
P_{1,b}&=A\abs{v}^2\int_1^{T} e^{-A\abs{v-\eta_1}^2 \left(\frac{2^R}{1+\tau^{1/\delta}}-1\right)^{\delta}}e^{-A\abs{v}^2\tau}~d\tau\nonumber\\
&\geq A\abs{v}^2\int_1^{T} e^{-A\abs{v-\eta_1}^2 \left(2^R/2-1\right)^{\delta}} e^{-A\abs{v}^2\tau}~d\tau\nonumber\\
&\overset{(e)}{=} e^{-A\abs{v-\eta_1}^2 \bar{T}} \left(e^{-A\abs{v}^2}-e^{-AT\abs{v}^2}\right) \label{II_pr}
\end{align}
where in (e) $\bar{T}=\left(2^R/2-1\right)^{\delta}$.
Combining (\ref{I_pr}) and (\ref{II_pr}), (\ref{Pr_in_te}) is rewritten as
\begin{align}
\mathbb{P} \left(I_1+I_2 \geq R\right)
&\geq e^{-A\abs{v-\eta_1}^2T} \left(1-e^{-A
\abs{v}^2}\right)
+e^{-A\abs{v-\eta_1}^2 \bar{T}} \left(e^{-A\abs{v}^2}-e^{-AT\abs{v}^2}\right)\nonumber\\
&+e^{-AT\abs{v}^2}
\label{P_1fin}
\end{align}

\subsection{Derivation of $\mathbb{E}\big[\abs{\mathrm{\Sigma}_2}\big]$}
\label{der_exp_irB}
From (\ref{are_eq}),
\begin{align}
\mathbb{E}\big[\abs{\mathrm{\Sigma}_2}\big]&=
\int_{\mathbb{R}^2}\mathbb{P}
\left(I_1+I_2 \geq R\right)~\ud v\nonumber\\
&\geq \int_{\mathbb{R}^2}\Big[ e^{-A\abs{v-\eta_1}^2T} \left(1-e^{-A\abs{v}^2}\right)+e^{-A\abs{v-\eta_1}^2 \bar{T}} \left(e^{-A\abs{v}^2}-e^{-AT\abs{v}^2}\right)+e^{-AT\abs{v}^2}\Big ] \ud v\nonumber\\
&=\int_{\mathbb{R}^2}\Big[e^{-A\abs{v-\eta_1}^2T}-
e^{-A\left(T\abs{v-\eta_1}^2+\abs{v}^2\right)}+ e^{-A\left(\bar{T}\abs{v-\eta_1}^2+\abs{v}^2\right)}-
e^{-A\left(\bar{T}\abs{v-\eta_1}^2+T\abs{v}^2\right)}\nonumber\\
&~~+e^{-AT\abs{v}^2}\Big ] \ud v\nonumber\\
&\equiv H_1-H_2+H_3-H_4+H_5
\label{are_eqII}
\end{align}
To evaluate the 5 integrals in (\ref{are_eqII}), we first write down the following general integral in simple form as
\begin{align}
H&=\int_{\mathbb{R}^2}e^{-A\left(c_1\abs{v}^2+c_2 \abs{v-\eta_1}^2\right)}~\ud v\nonumber\\
&=\iint e^{-A\left(c_1\left(v_1^2+v_2^2\right)+c_2\left(v_1-\tilde{d}_1\right)^2+c_2v_2^2\right)}~\ud v_1~\ud v_2\nonumber\\
&=\int_{-\infty}^{\infty}e^{-A\left(c_1v_1^2+c_2\left(v_1-\tilde{d}_1\right)^2\right)}~\ud v_1~\cdot~ \int_{-\infty}^{\infty}e^{-A\left(c_1+c_2\right) v_2^2}~\ud v_2
\label{ge_H_int}
\end{align}
The exponent in the $1^{st}$ integral of (\ref{ge_H_int}) is rewritten by completing squares as
\begin{equation}
c_1v_1^2+c_2\left(v_1-\tilde{d}_1\right)^2=\left(c_1+c_2\right) \left(v_1-\frac{\tilde{d}_1c_2}{c_1+c_2}\right)^2+\frac{\tilde{d}_1^2c_1c_2}
{c_1+c_2}\label{cs_rw}
\end{equation}
Using (\ref{cs_rw}), the integral $H$ in (\ref{ge_H_int}) becomes
\begin{align}
H&=e^{-\frac{A\tilde{d}_1^2c_1c_2}{c_1+c_2}}\int_{-\infty}^{\infty}
e^{-A\left(c_1+c_2\right)\left(v_1-\frac{\tilde{d}_1c_2}
 {c_1+c_2}\right)^2}~\ud v_1~\cdot~ \int_{-\infty}^{\infty} e^{-A\left(c_1+c_2\right) v_2^2}~\ud v_2\label{H_int}
\end{align}
To evaluate (\ref{H_int}), the standard Gaussian pdf relation is rewritten in the following manner
\begin{align}
&\int_{-\infty}^{\infty}\frac{1}{\sqrt{2\pi\sigma^2}}~e^{-\left( x-\mu\right)^2\big /2\sigma^2}~\ud x=1\nonumber\\
\Rightarrow &\int_{-\infty}^{\infty}e^{-b\left(x-\mu\right)^2}~\ud x=\sqrt{\frac{\pi}{b}}\label{ref_int_eq}
\end{align}
where $b=\frac{1}{2\sigma^2}$.
Applying the relation in (\ref{ref_int_eq}) to the integral $H$ in (\ref{H_int}) twice yields
\begin{align}
H&=\int_{\mathbb{R}^2}e^{-A\left(c_1\abs{v}^2+c_2 \abs{v-\eta_1}^2\right)}~\ud v\nonumber\\
&=e^{-\frac{A\tilde{d}_1^2c_1c_2}{c_1+c_2}}~\frac{\pi}
{A\left(c_1+c_2\right)}\label{Ref_it}
\end{align}

Now the integrals in (\ref{are_eqII}) can be evaluated using the result in (\ref{Ref_it}). The lower bound for $\mathbb{E}\big[\abs{\mathrm{\Sigma}_2}\big]$ is given as
\begin{align}
\mathbb{E}\big[\abs{\mathrm{\Sigma}_2}\big]
&\geq H_1-H_2+H_3-H_4+H_5\nonumber\\
&=\frac{\pi}{A}\Bigg[\frac{1}{T}-\frac{e^{-AT
\tilde{d}_1^2/\left(1+T\right)}} {1+T}+\frac{e^{-A\bar{T}\tilde{d}_1^2/\left(1+\bar{T}\right)}} {1+\bar{T}}-\frac{e^{-AT\bar{T}\tilde{d}_1^2/ \left(T+\bar{T}\right)}} {T+\bar{T}}+\frac{1}{T}
\Bigg]
\end{align} 

\section{Repetition Combining}
\label{der_exp_rtd}
As in section \ref{der_exp_ir}, we first obtain an expression for $\mathbb{P} \left(\mathrm{SIR}\left(v,0\right)+\mathrm{SIR}\left(v,\eta_1\right) \geq 2^R-1\right)$ and proceed to $\mathbb{E}\big[\abs{\mathrm{\Sigma}_2}\big]$.
Using the notation for $\mathrm{SIR}\left(v,0\right)$ in (\ref{SIR_not}), we have
\begin{align}
&\mathbb{P} \left(\mathrm{SIR}\left(v,0\right)+\mathrm{SIR}\left(v,\eta_1\right) \geq 2^R-1\right)\nonumber\\
&=\mathbb{P}\left(S_1\abs{v}^{-\alpha}+S_2\abs{v-\eta_1}^{-\alpha}
\geq T^{\frac{1}{\delta}}\right)\nonumber\\
&=\int_0^{\infty}\mathbb{P}\left(S_2\geq \abs{v-\eta_1}^{\alpha} \left(T^{\frac{1}{\delta}}-s_1\abs{v}^{-\alpha}\right)\right)
f_{S_1}\left(s_1\right)\mathrm{d}s_1\nonumber\\
&\overset{(a)}{=}\int_0^{T^{\frac{1}{\delta}}\abs{v}^{\alpha}} e^{-A\abs{v-\eta_1}^2\left(T^{\frac{1}{\delta}}-s_1 \abs{v}^{-\alpha}\right)^{\delta}} f_{S_1}\left(s_1\right) \mathrm{d}s_1
+\int_{T^{\frac{1}{\delta}}\abs{v}^{\alpha}}^{\infty} f_{S_1}\left(s_1\right) \mathrm{d}s_1\nonumber\\
&=\int_0^{T^{\frac{1}{\delta}}\abs{v}^{\alpha}}e^{-A\abs{v-\eta_1}^2
\left(T^{\frac{1}{\delta}}-s_1\abs{v}^{-\alpha}\right)^{\delta}} e^{-As_1^{\delta}}~A~\ud\left(s_1^{\delta}\right)
+e^{-AT\abs{v}^2}\nonumber\\
&\overset{(b)}{=} \int_0^T e^{-A\abs{v-\eta_1}^2\left(T^{\frac{1}{\delta}} -u^{1/{\delta}}\right)^{\delta}}e^{-A\abs{v}^2u}
~A\abs{v}^2~\mathrm{d}u
+e^{-AT\abs{v}^2}\nonumber\\
&\overset{(c)}{\geq}e^{-AT\abs{v}^2}+ e^{-AT\abs{v-\eta_1}^2}\int_0^1
A\abs{v}^2~e^{-A\abs{v}^2u} \mathrm{d}u\nonumber\\
&~~+ e^{-A\abs{v-\eta_1}^2\left(T^{\frac{1}{\delta}} -1\right)^{\delta}}\int_1^T A\abs{v}^2~e^{-A\abs{v}^2u} \mathrm{d}u
\nonumber\\
&\overset{(d)}{\geq}e^{-AT\abs{v}^2}+e^{-AT\abs{v-\eta_1}^2}\left(1-e^{-A\abs{v}^2}\right)+
e^{-A\tilde{T}\abs{v-\eta_1}^2}\left(e^{-A\abs{v}^2}-e^{-AT\abs{v}^2}\right)
\label{P_SIR_exp}
\end{align}
where (a) uses the CCDF of $S_2$ in (\ref{S_ccdf}), (b) follows from the previous line using the simple substitution $u=s_1^{\delta}\abs{v}^{-2}$, using the same steps as in Appendix \ref{der_exp_ir} after (\ref{P_1equ}) results in (c) and in (d), we use $\tilde{T}=\left(2^R-2\right)^{\delta}$.

Now (\ref{P_SIR_exp}) is similar to (\ref{P_1fin}) with $\tilde{T}$ replacing $\bar{T}$ and the rest of the proof follows the same steps as in Appendix \ref{der_exp_irB}.

\bibliography{References_II}
\bibliographystyle{IEEEtran}
\end{document}